\documentclass{article}
\newcommand{\Dzero}{D\O}
\newcommand{\runi}{Run~I}
\newcommand{\runii}{Run~II}

\newcommand{\W}{\ensuremath{W}}
\newcommand{\Z}{\ensuremath{Z}}
\newcommand{\Hp}{\ensuremath{H^+}}
\newcommand{\bquark}{\ensuremath{b}}
\newcommand{\q}{\ensuremath{q}}
\newcommand{\topquark}{\ensuremath{t}}
\newcommand{\tprime}{\topquark'}
\newcommand{\bbbar}{\ensuremath{\bquark{\overline{\bquark}}}}
\newcommand{\enu}{\ensuremath{e\nu}}
\newcommand{\emu}{\ensuremath{e\mu}}
\newcommand{\munu}{\ensuremath{\mu\nu}}
\newcommand{\taunu}{\ensuremath{\tau\nu}}

\newcommand{\ppbar}{\ensuremath{p\bar{p}}}
\newcommand{\ttbar}{\ensuremath{t\bar{t}}}

\newcommand{\ejets}{e+jets}
\newcommand{\mujets}{\ensuremath{\mu}+jets}

\newcommand{\Vtb}{\ensuremath{V_{\topquark\bquark}}}

\newcommand{\mtop}{\ensuremath{m_{top}}}

\newcommand{\mW}{\ensuremath{m_{W}}}

\newcommand{\Et}{\ensuremath{E_{T}}}
\newcommand{\Etmiss}{\ensuremath{E \kern-0.6em\slash_{T}}}
\newcommand{\Etmissx}{\ensuremath{E \kern-0.6em\slash_{x}}}
\newcommand{\Etmissy}{\ensuremath{E \kern-0.6em\slash_{y}}}

\newcommand{\pt}{\ensuremath{p_{T}}}
\newcommand{\Ht}{\ensuremath{H_{T}}}

\newcommand{\sigmattbar}{\ensuremath{\sigma(\ttbar)}}
\newcommand{\pandm}[2]{\ensuremath{^{+{#1}}_{-{#2}}}}
\newcommand{\stat}{\ensuremath{({\rm stat.})}}
\newcommand{\syst}{\ensuremath{({\rm syst.})}}
\newcommand{\lumi}{\ensuremath{({\rm lumi.})}}

\newcommand{\pb}{\,\ensuremath{\rm pb}}
\newcommand{\invpb}{\,\ensuremath{\rm pb^{-1}}}
\newcommand{\MeV}{\,\ensuremath{\mathrm{Me\kern-0.1em V}}}
\newcommand{\GeV}{\,\ensuremath{\mathrm{Ge\kern-0.1em V}}}
\newcommand{\TeV}{\,\ensuremath{\mathrm{Te\kern-0.1em V}}}

\usepackage{LaThuileFPSpro}
\begin{document}
\title{ 
  TOP QUARK PRODUCTION AND PROPERTIES
  AT THE TEVATRON
  }
\author{
  Frank Fiedler\\
  {\em Munich University (LMU), Germany}\\
  on behalf of the CDF and \Dzero\ Collaborations
  }
\maketitle

\baselineskip=11.6pt

\begin{abstract}
  The precise measurement of top quark production and properties is
  one of the primary goals of the Tevatron during \runii.  
  The total $\ttbar$ production cross-section has been measured in 
  a large variety of decay channels and using different selection 
  criteria.  Results from differential cross-section measurements and
  searches for new physics in $\ttbar$ production and top quark 
  decays are available.  
  Electroweak production of single top quarks has been searched for.
  The results from all
  these analyses, using typically $200\invpb$ of data, are presented.
\end{abstract}
\newpage
\section{Introduction}

The top quark is special among the fermions of the Standard Model
because of its large mass.
Currently, the top quark can only be studied at the two Tevatron 
experiments CDF and \Dzero, where
measurements of top quark production and properties
are one of the key physics goals of \runii.

The top quark mass
is discussed in a separate article\cite{bib-velev}.
This article focuses on measurements of the total
$\ttbar$ production cross-section, 
searches for new physics in $\ttbar$ production and top quark decay,
and on the search for electroweak (single) 
top quark production.
While the CDF and \Dzero\ experiments have both collected more than
$500\invpb$ of data so far during Tevatron \runii, surpassing the 
\runi\ integrated luminosity by a factor $\ge$5, the measurements
summarized in this article typically use about $200\invpb$.

In section~\ref{topproduction.sec}, general aspects of
top production and event
topologies at the Tevatron are briefly discussed.
Section~\ref{ttbarxs.sec} discusses the measurements of the 
total $\ttbar$ production cross-section,
while further measurements in $\ttbar$ events are 
presented in section~\ref{furtherttbar.sec}.
The search for single top quark production is 
presented in section~\ref{singletop.sec}.

\section{Top Quark Production at the Tevatron}

\label{topproduction.sec}

\begin{figure}[t]
  \vspace{5.4cm}
  \includegraphics{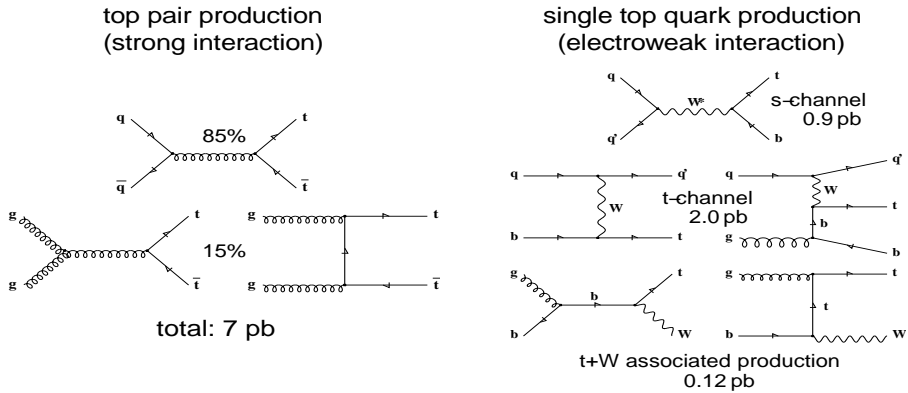}
  \caption{\it
    Leading order Feynman diagrams for top quark production at
    hadron colliders together with the Standard Model cross-sections
    at the Tevatron.
    \label{feyndiag.fig} }
\end{figure}

In the Standard Model, the
production of top quarks at a hadron collider can in principle
proceed via two mechanisms: 
$\ttbar$ pair production via the 
strong interaction, and single top (or antitop) production via
the electroweak interaction.
The leading order Feynman diagrams are shown in
figure~\ref{feyndiag.fig} together with the Standard Model
cross-sections in $\ppbar$ collisions at a centre-of-mass energy
of 1.96~TeV (corresponding to the Tevatron collider 
at \runii)\cite{bib-theoryttbarxs,bib-theorysingletopxs}.

In the following, the main characteristics of $\ttbar$ and single
top events at the Tevatron are discussed.

\subsection{Classification of $\ttbar$ Event Topologies}

In the Standard Model, the branching fraction of top quark decays to
a b quark and an on-shell \W\ boson is close to $100\%$, other decay
modes not being observable with Tevatron luminosities.
The subsequent \W\ decays determine the event topology seen
in the detector, and 
$\ttbar$ events are classified as follows:
\begin{list}{$\bullet$}{\setlength{\itemsep}{0ex}
                        \setlength{\parsep}{0ex}
                        \setlength{\topsep}{0ex}}
\item
{\bf Dilepton events}, where both \W\ bosons decay into an $\enu$ or 
$\munu$ final state, are characterized by two energetic, isolated
leptons of opposite charge, two energetic b jets, and missing 
transverse energy. While the product branching ratio is only about
$5\%$, pure event samples can be obtained requiring the two leptons
in the event to be reconstructed.
\item
In {\bf lepton+jets events}, one \W\ boson decays hadronically and the
other into an $\enu$ or $\munu$ final state.  This topology is 
characterized by an energetic, isolated electron or muon, four
energetic jets (two b jets and two light-quark jets from the \W\ decay),
and missing transverse energy.  The product branching ratio of
$\approx\!30\%$ is larger than for dilepton events, and
the main background is from $\W$+jets events.
\item
In {\bf hadronic events}, one expects 6 energetic jets (of which two
are b jets) and no significant missing transverse energy.  Because of 
large backgrounds from QCD jet production, identifying $\ttbar$ events
in the hadronic channel is challenging, despite the large 
product branching ratio of $\approx\!44\%$.
\item
In about $21\%$ of the $\ttbar$ events, at least one \W\ boson decays
into a {\bf\boldmath$\taunu$\unboldmath{} final state}.  Depending on its decay, the $\tau$
lepton can be identified as a narrow jet, an isolated track, or an electron or 
muon.  Two energetic b jets, missing transverse energy, and the decay
products from the second \W\ boson complete the event topology.
\end{list}

In general, the reconstruction and selection of $\ttbar$ event
candidates is based on reconstructing the directions and
energies/momenta of isolated electrons or muons and jets, and on 
reconstructing the missing transverse energy $\Etmiss$ from the 
transverse momentum balance in the event.
The purity of the event samples can be enhanced by
identifying jets that originated from a b quark (b tagging), since in
the Standard Model, every $\ttbar$ event contains two b jets.
Both CDF and \Dzero\ use
\begin{list}{$\bullet$}{\setlength{\itemsep}{0ex}
                        \setlength{\parsep}{0ex}
                        \setlength{\topsep}{0ex}}
\item
secondary vertex algorithms, based on explicit reconstruction of
the decay vertex of the b hadron within the jet; 
\item
impact parameter based algorithms that classify tracks inside a jet
according to their distance of closest approach to the primary event 
vertex; and
\item
soft leptons from semileptonic bottom or
charm hadron decay (only muons are used so far)
\end{list}
to identify b jets.
The requirements on the jet multiplicity, the minimum jet
transverse energies, b identification of the jets, and event kinematic
information can be balanced to minimize the measurement error;
depending on the selection, not all jets need to be explicitly 
reconstructed.

\subsection{Single Top Quark Production}

\begin{figure}[t]
  \vspace{3.0cm}
  \includegraphics{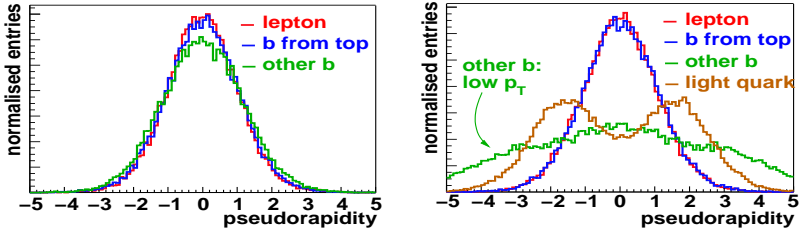}
  \caption{\it
    Expected pseudorapidity distributions of the charged lepton and 
    jets in single top events (left: s-channel, right: t-channel)\protect\cite{bib-dzerosingletop}.
    \label{singletopdistributions.fig} }
\end{figure}

The total cross-section for single top quark production is only a 
factor $\sim$2 smaller than that for $\ttbar$ production; however, 
the relevant backgrounds are substantially larger (\W+2jet 
instead of \W+4jet events).
To reduce the background, the 
selection of single top event candidates focuses on top decays
with leptonic $\W$ decays and on the identification of the b jet(s)
in the event.
Figure~\ref{singletopdistributions.fig} shows the expected 
pseudorapidity distributions of the charged lepton and jets in 
single top events.
For s-channel events, two b jets are expected in the center 
of the detector.  In general only one b jet can be reconstructed in 
case of the t-channel, but here an additional light quark jet can 
be observed.

\section{Measurements of the Total \boldmath$\ttbar$\unboldmath\ Production Cross-Section}

\label{ttbarxs.sec}

The goal is to measure the $\ttbar$ cross-section in as many different
modes as possible to check the predictions of the Standard Model.
The measurements in the different channels are described in the
following sub-sections.

\subsection{Lepton+Jets Channel, Topological Analyses}
\label{ljetstopoxs.sec}

\begin{figure}[t]
  \vspace{5.2cm}
  \includegraphics{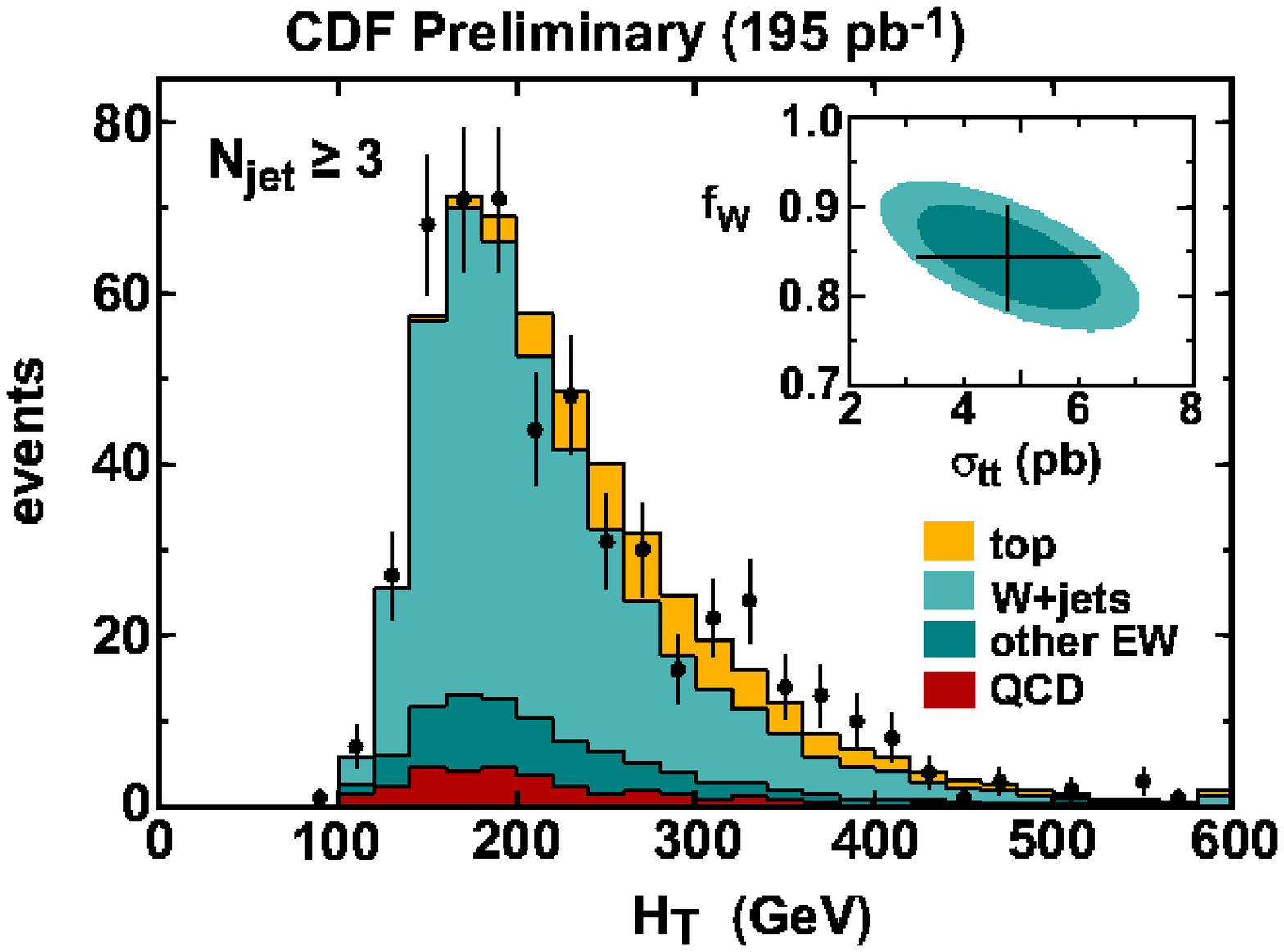}
  \includegraphics{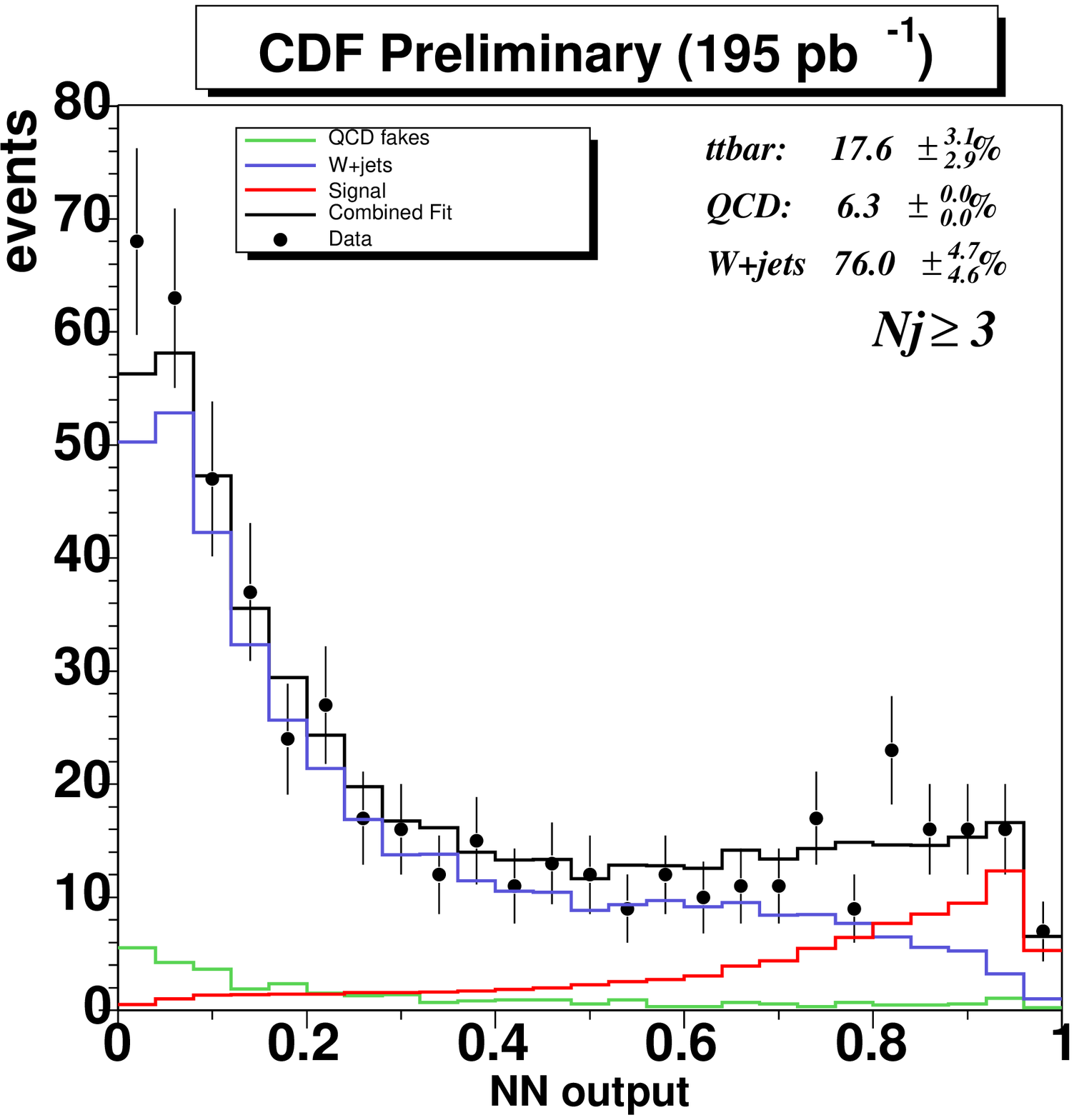}
  \caption{\it
    CDF $\ttbar$ cross-section measurement in the lepton+jets channel
    using topological information.  
    Left: $\Ht$ distribution.  Right: NN output.
    \label{CDFljetstopoxs.fig} }
\end{figure}

Both CDF and \Dzero\ have measured the $\ttbar$ cross-section
in the lepton+jets channel without relying on b jet identification.
In the CDF analysis\cite{bib-CDFljetstopoxs}, events with one
isolated electron with $\Et>20\GeV$ or muon with $\pt>20\GeV$, 
missing transverse energy
$\Etmiss>20\GeV$, and at least 3 jets with $\Et>15\GeV$ within
the pseudorapidity range $|\eta|<2.0$ are selected.
For $\Etmiss<30\GeV$, there is an additional cut requiring 
the angle $\Delta\phi$ between the missing transverse energy and the highest
$\Et$ jet in the transverse plane to satisfy $0.5<\Delta\phi<2.5$.
After this selection, the $\ttbar$ signal can be seen in the 
$\Ht$ distribution ($\Ht$ is the scalar sum of transverse energies of
the lepton, jets, and the missing transverse energy) 
as shown in figure~\ref{CDFljetstopoxs.fig}, and the
$\ttbar$ cross-section is measured to be
\begin{equation}
\sigmattbar = \left(4.7 \pm 1.6\stat \pm 1.8\syst\right)\pb 
\end{equation}
using a $195\invpb$ data sample.
To optimise the measurement, 7 quantities have been 
chosen for training a neural network (NN) 
to separate the $\ttbar$ signal from the background.
The NN output distribution is also shown in
figure~\ref{CDFljetstopoxs.fig}.
From a fit to this distribution a result of
\begin{equation}
\sigmattbar = \left(6.7 \pm 1.1\stat \pm 1.6\syst\right)\pb 
\end{equation}
is obtained from the same dataset.
In both analyses, the main systematic error is from the uncertainty in
the jet energy scale; it is however reduced from $30\%$ in the 
$\Ht$ based measurement to $16\%$ in the optimised analysis.

\begin{figure}[t]
  \vspace{3.7cm}
  \includegraphics{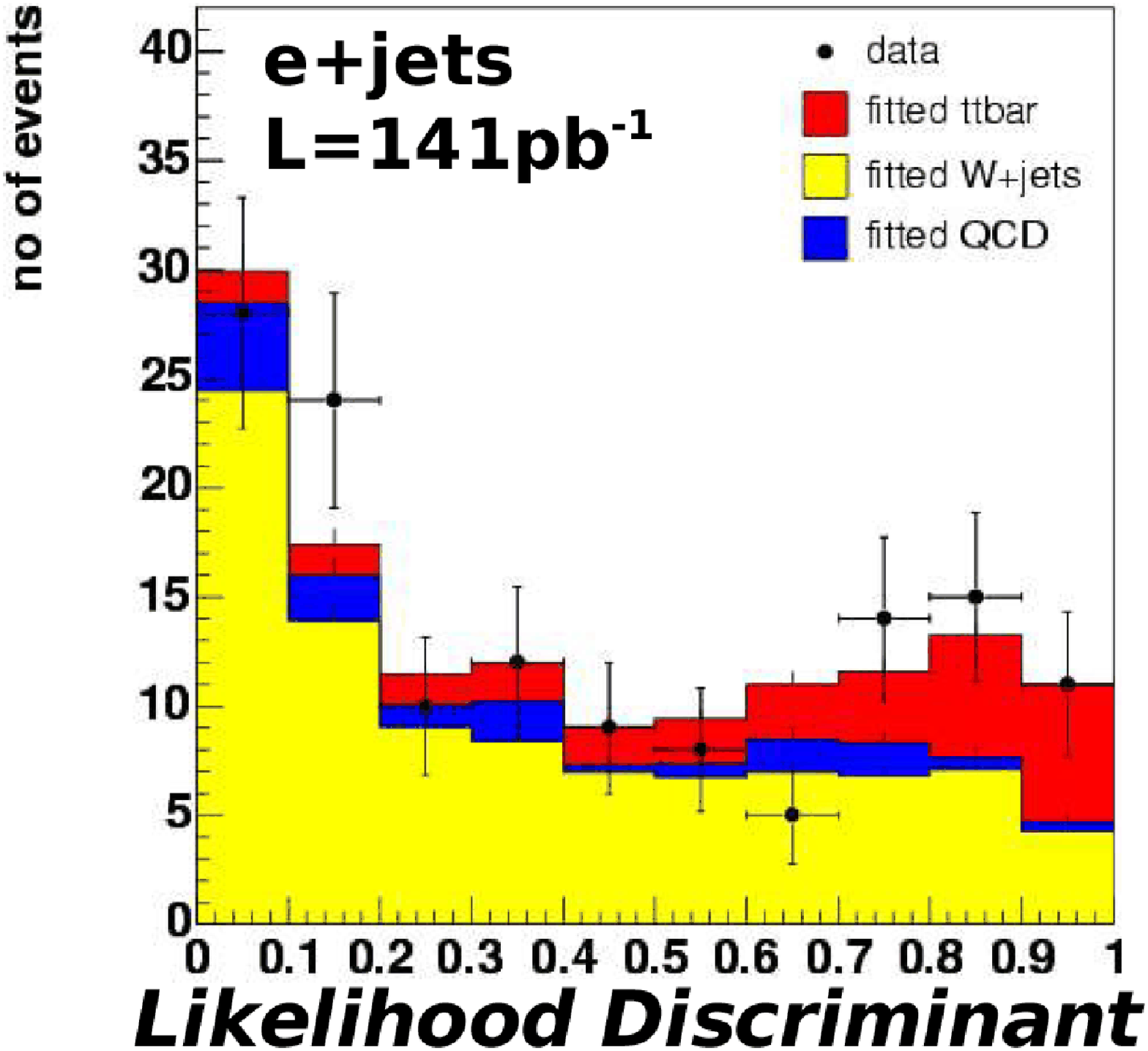}
  \includegraphics{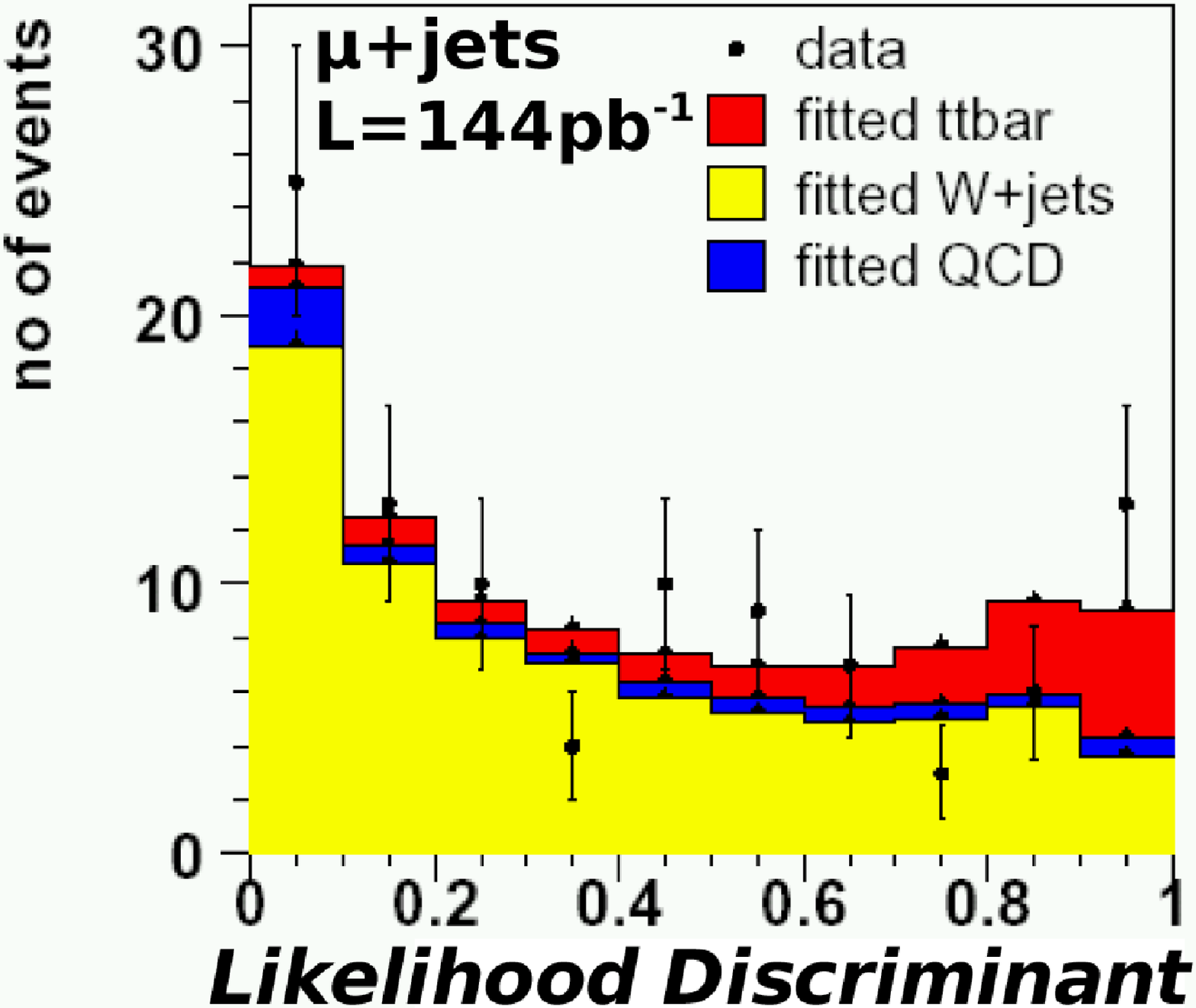}
  \caption{\it
    \Dzero\ $\ttbar$ cross-section measurement in the lepton+jets channel
    using topological information.  
    The likelihood discriminant distributions are shown for \ejets\
    (left) and \mujets\ (right) events together with the fitted
    $\ttbar$ and background contributions.
    \label{Dzeroljetstopoxs.fig} }
\end{figure}

In the \Dzero\ topological analysis\cite{bib-dzeroljetstopoxs}, 
events with one 
isolated electron or muon with $\pt>20\GeV$, missing transverse
energy ($\Etmiss>20\,\GeV$ in the \ejets\ case and $\Etmiss>17\GeV$
for \mujets\ events), and at least four jets with $\Et>15\GeV$ within
$|\eta|<2.5$ are selected.
So in contrast to CDF, four jets are required to be reconstructed, but
a larger pseudorapidity region is allowed.
To further separate $\ttbar$ events from background, a likelihood is 
constructed using angular variables and ratios of
energy dependent variables, to avoid direct dependence on the jet
energy scale.
The resulting distributions are given 
in figure~\ref{Dzeroljetstopoxs.fig}.
The combined fit to the \ejets\ and \mujets\ distributions
from $141-144\invpb$ of data yields
\begin{equation}
\sigmattbar = \left(7.2 \pandm{2.6}{2.4}\stat 
                        \pandm{1.6}{1.7}\syst 
                        \pm 0.5\lumi\right)\pb \ .
\end{equation}

\subsection{Lepton+Jets Channel, b Tagging Analyses}

\begin{figure}[t]
  \vspace{4.5cm}
  \includegraphics{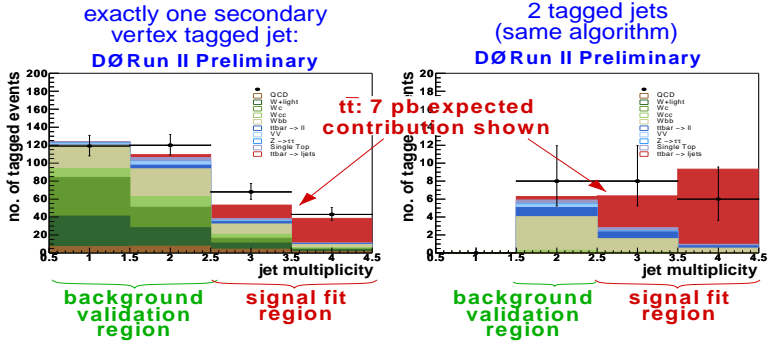}
  \caption{\it
    \Dzero\ $\ttbar$ cross-section measurement in the lepton+jets channel
    using b tagging information.  
    The jet multiplicity distributions for single and double secondary
    vertex tagged events are shown together with the expected Standard
    Model signal and background.
    \label{Dzeroljetsbtagxs.fig} }
\end{figure}

For the \Dzero\ measurements that make use of lifetime b tagging 
information\cite{bib-dzeroljetsbtagxs}, 
events are selected with the same criteria as above.
The $\ttbar$ cross-section is then determined from a combined fit
to the jet multiplicity distributions for events with exactly one 
b tagged jet and events with at least two b tagged jets.
Using a data sample of $158-169\invpb$, \Dzero\ obtains
\begin{equation}
\sigmattbar = \left(8.2 \pm 1.3\stat
                        \pandm{1.9}{1.6}\syst 
                        \pm 0.5\lumi\right)\pb
\end{equation}
using secondary vertex b tagging and
\begin{equation}
\sigmattbar = \left(7.2 \pandm{1.3}{1.2}\stat
                        \pandm{1.9}{1.4}\syst 
                        \pm 0.5\lumi\right)\pb
\end{equation}
with a track impact parameter based algorithm.
The jet multiplicity distributions obtained with secondary vertex
b tagging are shown in figure~\ref{Dzeroljetsbtagxs.fig}.
In a separate analysis, \Dzero\ analyzes events with a 
semimuonic bottom or charm decay, resulting in\cite{bib-Dzeroljetssoftmuonxs}
\begin{equation}
\sigmattbar = \left(11.2 \pm 4.0 \stat
                         \pm 1.3 \syst
                         \pm 1.1 \lumi\right)\pb
\end{equation}
based on $93\invpb$.

\begin{figure}[t]
  \vspace{4.5cm}
  \includegraphics{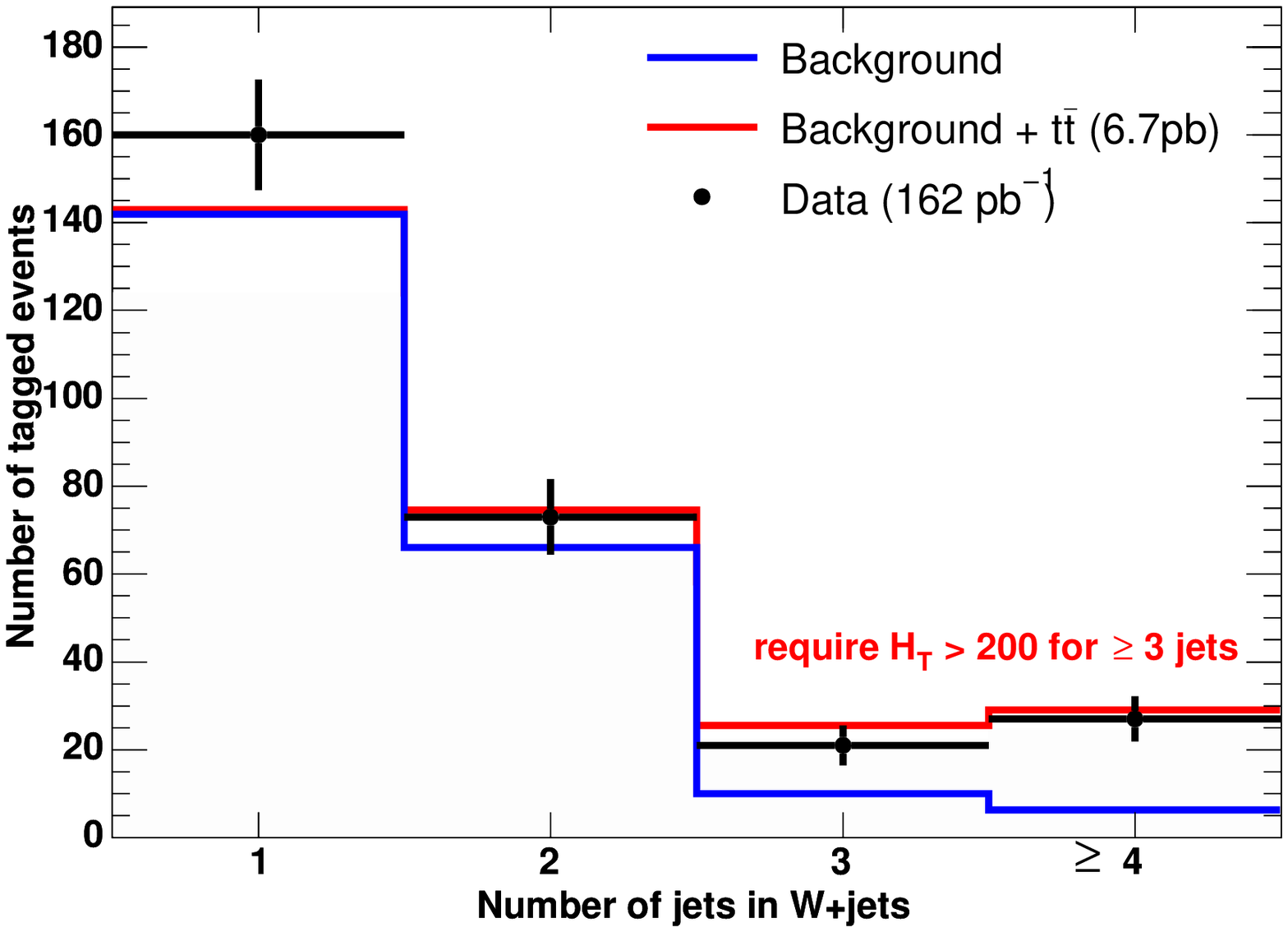}
  \includegraphics{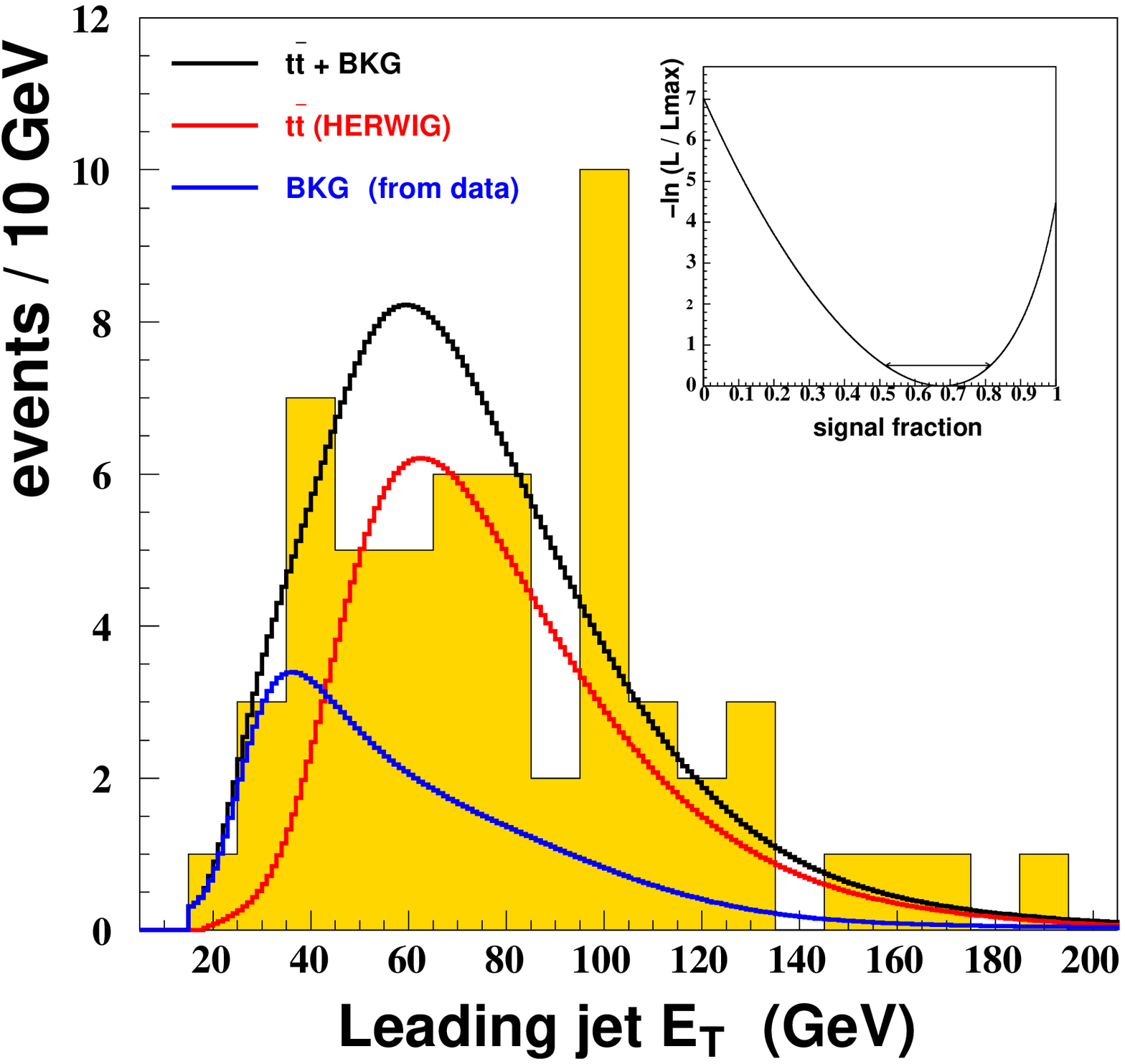}
  \caption{\it
    CDF $\ttbar$ cross-section measurement in the lepton+jets channel
    using b tagging information.  Left: the jet multiplicity
    distribution for events with at least one b tagged jet.  Right:
    the leading jet $\Et$ distribution for events with at least 3 jets.
    \label{CDFljetsbtagxs.fig} }
\end{figure}

Several CDF analyses make use of b tagging information.
The preselection of events requires one lepton, missing 
transverse energy, and three jets as in section~\ref{ljetstopoxs.sec}.
When at least one jet is required to be secondary vertex 
tagged\cite{bib-CDFljetsbtagxsanalysisone},
a measurement of
\begin{equation}
\sigmattbar = \left(5.6 \pandm{1.2}{1.1}\stat
                        \pandm{0.9}{0.6}\syst\right)\pb
\end{equation}
is obtained 
from the jet multiplicity distribution (where a value of $\Ht>200\GeV$
is required for events with three or more jets) in $162\invpb$
shown in figure~\ref{CDFljetsbtagxs.fig}.
Alternatively, the fraction of $\ttbar$ events in events
with at least 3 jets
is obtained from a fit to the $\Et$ distribution of the leading
jet, which is also shown in figure~\ref{CDFljetsbtagxs.fig},
yielding\cite{bib-CDFljetsbtagxsanalysistwo}
\begin{equation}
\sigmattbar = \left(6.0 \pm 1.6\stat \pm 1.2\syst\right)\pb \ .
\end{equation}
For the CDF multiple tag analysis, a special version of the b tagging 
algorithm has been developed with looser criteria to increase the 
statistics.
From the jet multiplicity distributions obtained with the regular
and the loose b tag, measurements of\cite{bib-CDFljetsbtagxsanalysisone,bib-CDFljetsloosedoublebtagxs}
\begin{eqnarray}
\sigmattbar = \left(5.0 \pandm{2.4}{1.9}\stat
                        \pandm{1.1}{0.8}\syst\right)\pb
&&
\ {\rm (regular\ b\ tag)\ and}
\\
\sigmattbar = \left(8.2 \pandm{2.4}{2.1}\stat
                        \pandm{1.8}{1.0}\syst\right)\pb
&&
\ {\rm (loose\ b\ tag)}
\end{eqnarray}
are obtained.
Finally, CDF also uses events with b jets identified by an
impact parameter based algorithm, yielding\cite{bib-CDFljetsimpparxs}
\begin{equation}
\sigmattbar = \left(5.8 \pandm{1.3}{1.2}\stat
                        \pm 1.3\syst\right)\pb
\end{equation}
in $162\invpb$, as well as
events with a semimuonic bottom or charm hadron
decay, resulting in\cite{bib-CDFljetssoftmuonxs}
\begin{equation}
\sigmattbar = \left(5.2 \pandm{2.9}{1.9}\stat
                        \pandm{1.3}{1.0}\syst\right)\pb
\end{equation}
using $200\invpb$.

\subsection{Dilepton Channel}

\begin{figure}[t]
  \vspace{3.7cm}
  \includegraphics{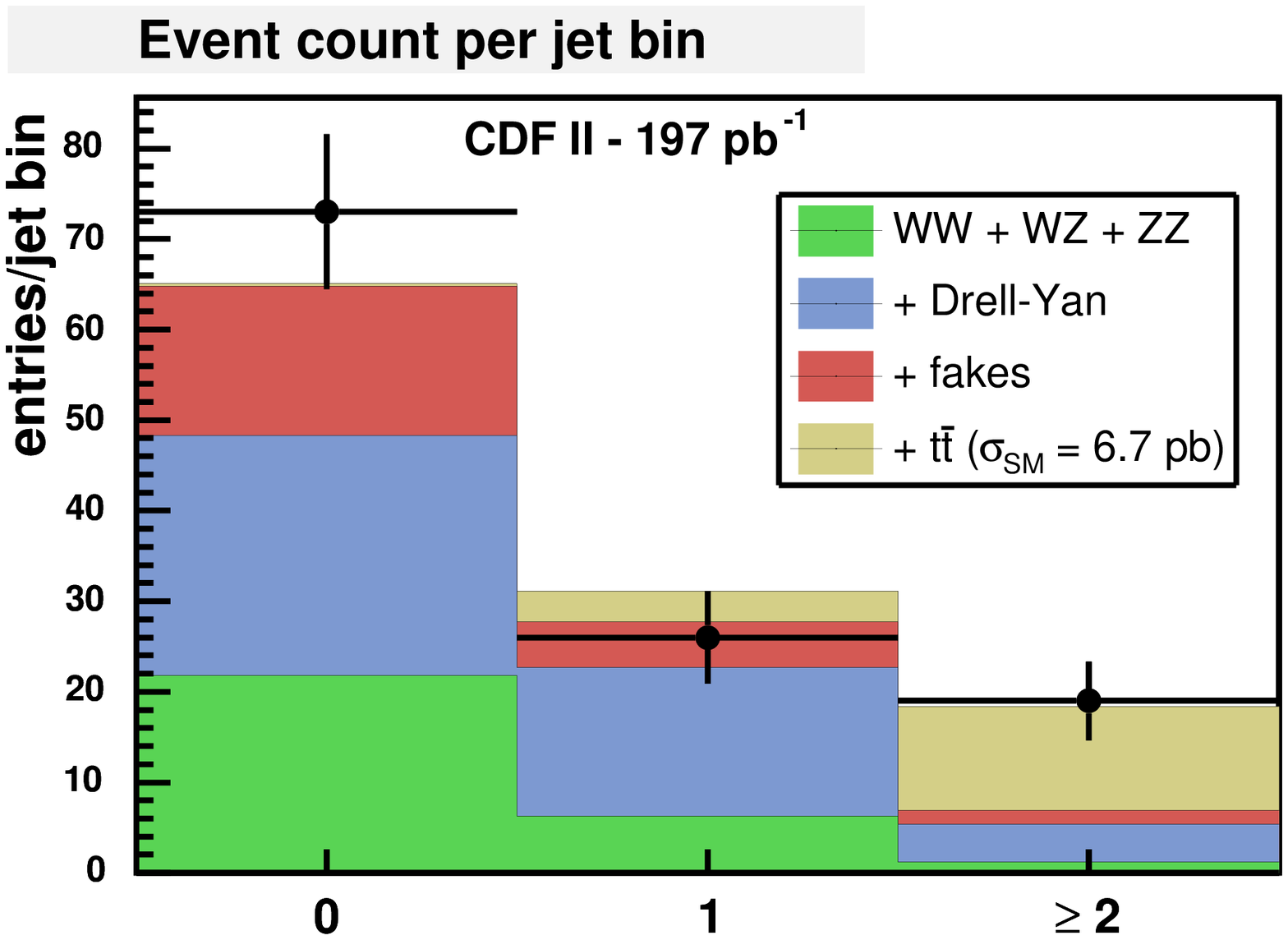}
  \includegraphics{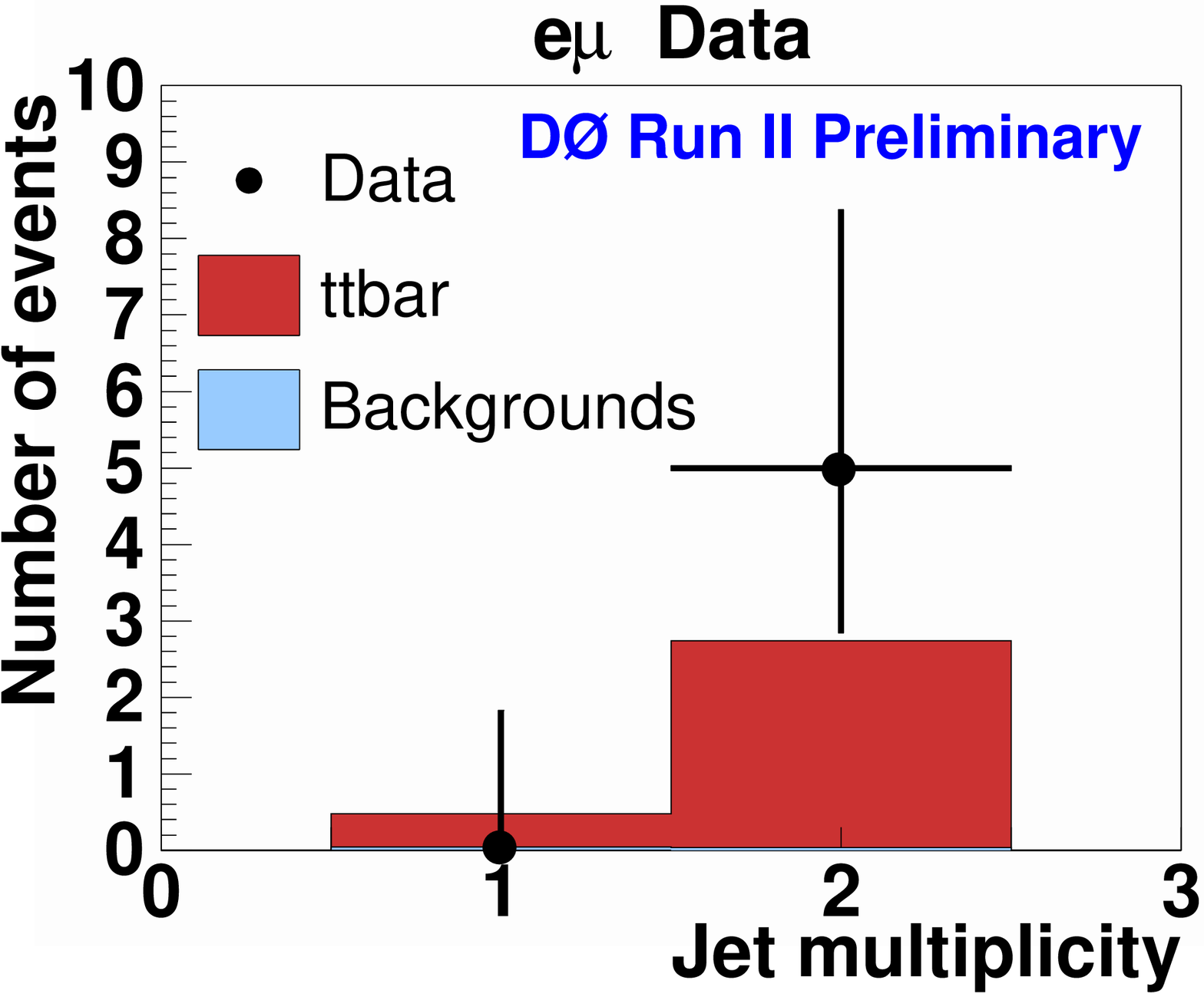}
  \includegraphics{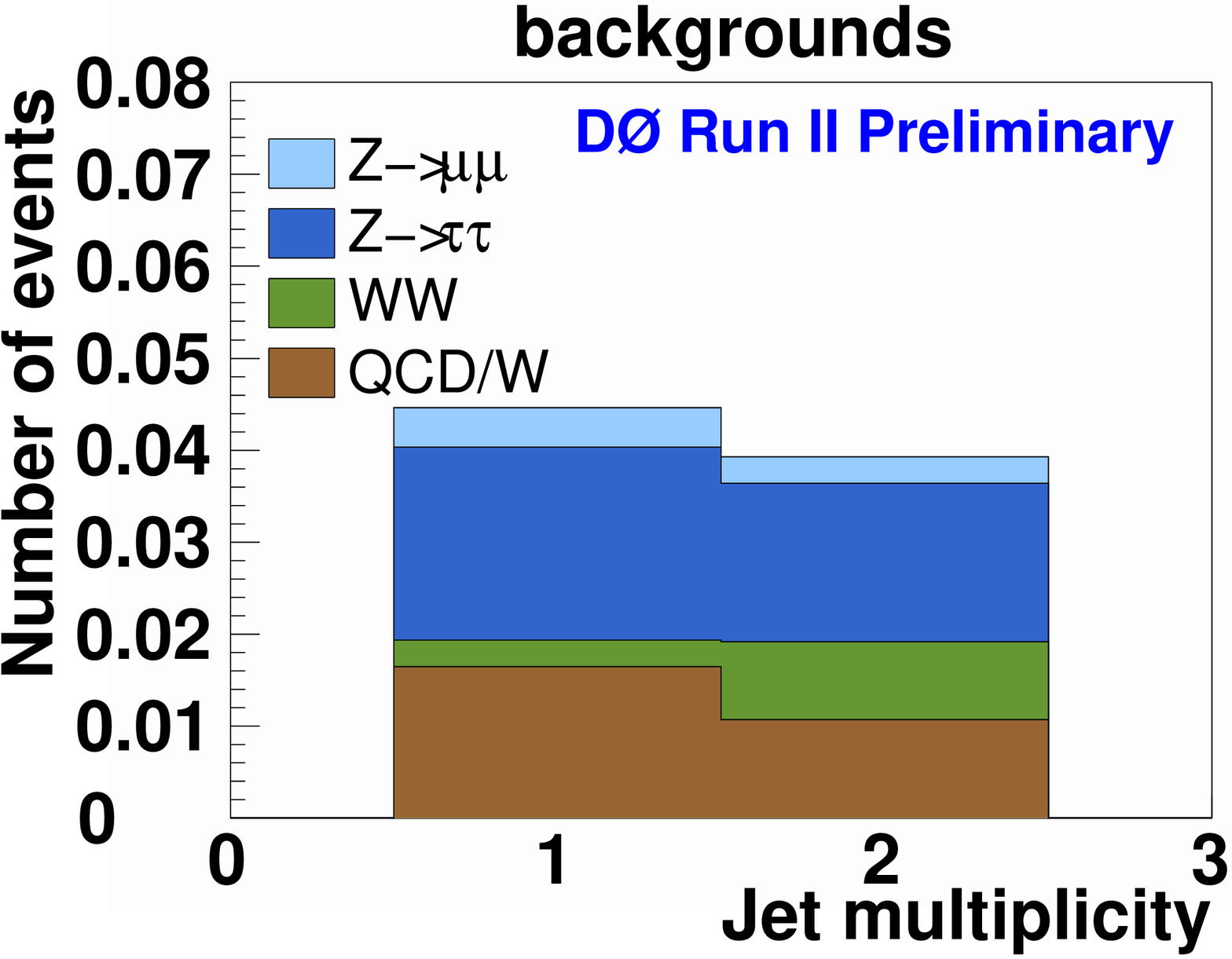}
  \caption{\it
    CDF and \Dzero\ $\ttbar$ cross-section measurements in the
    dilepton channel.  Left: jet multiplicity distribution in the 
    CDF ``lepton+track'' analysis.  Middle and right: jet multiplicity
    distributions for data (middle) and expected background (right) in 
    the \Dzero\ dilepton measurement with at least one secondary
    vertex b tagged jet.
    \label{dileptonxs.fig} }
\end{figure}

The CDF selection of $\ttbar$ events in the dilepton channel
requires two isolated tracks ($\pt>20\GeV$) and missing transverse
energy ($\Etmiss>25\GeV$).
The $\ttbar$ production cross-section is determined from the jet
multiplicity distribution of events where both tracks are identified
as leptons, or events where only one identified lepton is required.
The combined result is\cite{bib-CDFdileptonxsone}
\begin{equation}
\sigmattbar = \left(7.0 \pandm{2.4}{2.1}\stat
                        \pandm{1.6}{1.1}\syst
                        \pm 0.4\lumi \right)\pb
\end{equation}
using $200\invpb$.
The jet multiplicity distribution for the analysis with at least one identified
lepton is shown in figure~\ref{dileptonxs.fig}.
One of the main backgrounds to $\ttbar$ production in the dilepton 
channel is diboson (mostly \W\W) production.
In a separate analysis, CDF fits the two-dimensional jet multiplicity
vs.~$\Etmiss$ distribution to measure the $\ttbar$, $\W\W$, and 
$\Z\to\tau\tau$ cross-sections simultaneously.
This analysis yields\cite{bib-CDFdileptonxstwo}
\begin{equation}
\sigmattbar = \left(8.6 \pandm{2.5}{2.4}\stat
                        \pm 1.1\syst\right)\pb
\end{equation}

For the \Dzero\ dilepton analysis, events with two isolated leptons
with $\pt>15\GeV$ ($\pt>20\GeV$ in the dielectron channel), two jets with
$\Et>20\GeV$, missing transverse energy $\Etmiss>35\GeV$ 
($\Etmiss>25\GeV$ in the $\emu$ channel), and 
$\Ht^{\rm lead.\ \ell}>120 (140)\GeV$ in the $\mu\mu$ ($\emu$) channel
are selected, where $\Ht^{\rm lead.\ \ell}$ includes all jets and the 
leading lepton.
Additional cuts reject events consistent with a $\Z\to\ell\ell$
hypothesis.
With no b tagging criteria applied, the analysis yields\cite{bib-Dzerodileptontopoxs}
\begin{equation}
\sigmattbar = \left(14.3\pandm{5.1}{4.3}\stat
                        \pandm{2.6}{1.9}\syst
                        \pm 0.9\lumi\right)\pb
\end{equation}
using $140-156\invpb$.
When requiring at least one jet to be secondary vertex b tagged, 
the $\emu$ channel alone yields\cite{bib-Dzerodileptonbtagxs}
\begin{equation}
\sigmattbar = \left(11.1\pandm{5.8}{4.3}\stat
                        \pm 1.4\syst
                        \pm 0.7\lumi\right)\pb
\end{equation}
with a very high purity sample, see figure~\ref{dileptonxs.fig}.

\subsection{Hadronic Channel}

\begin{figure}[t]
  \vspace{4.4cm}
  \includegraphics{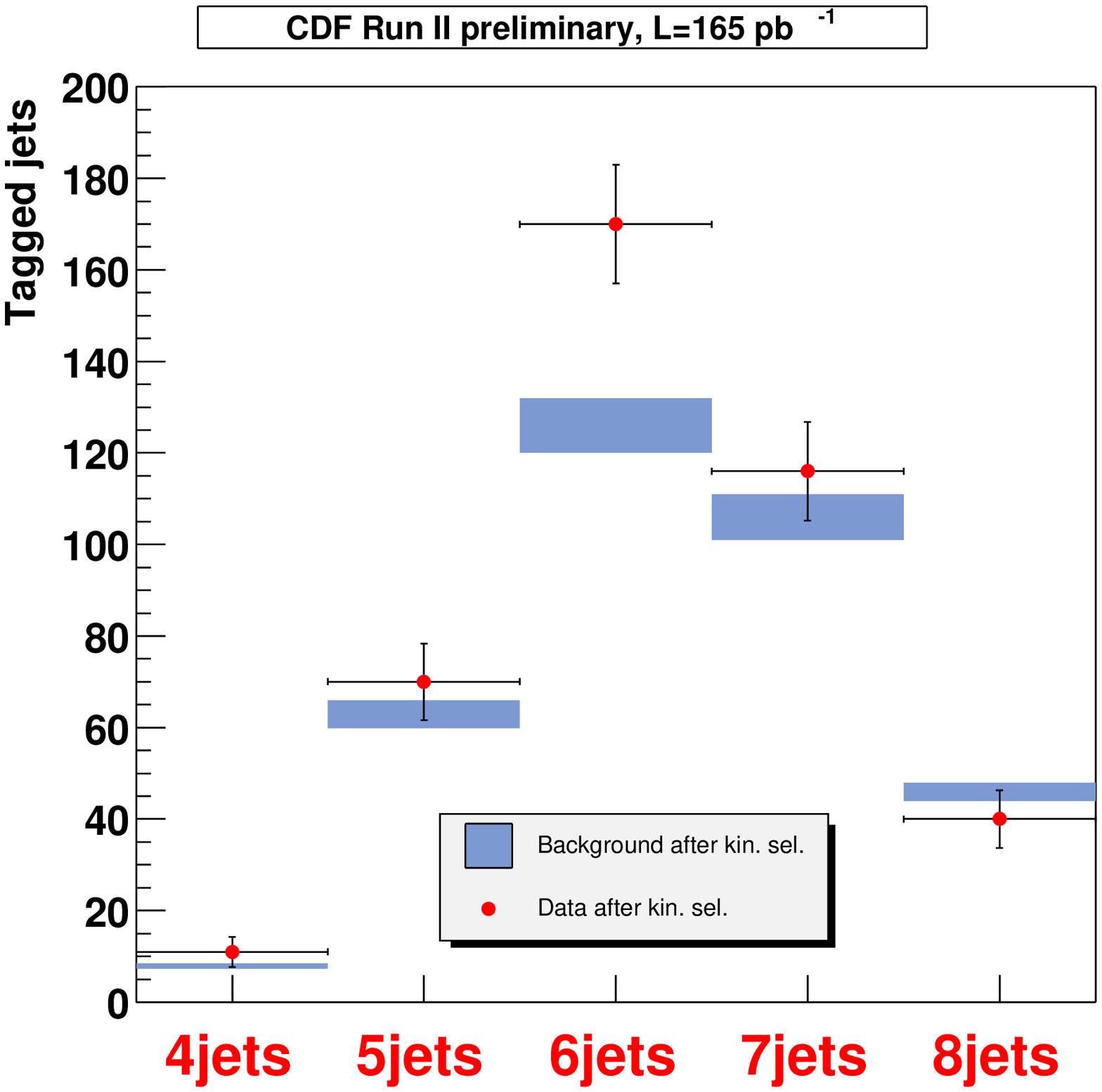}
  \includegraphics{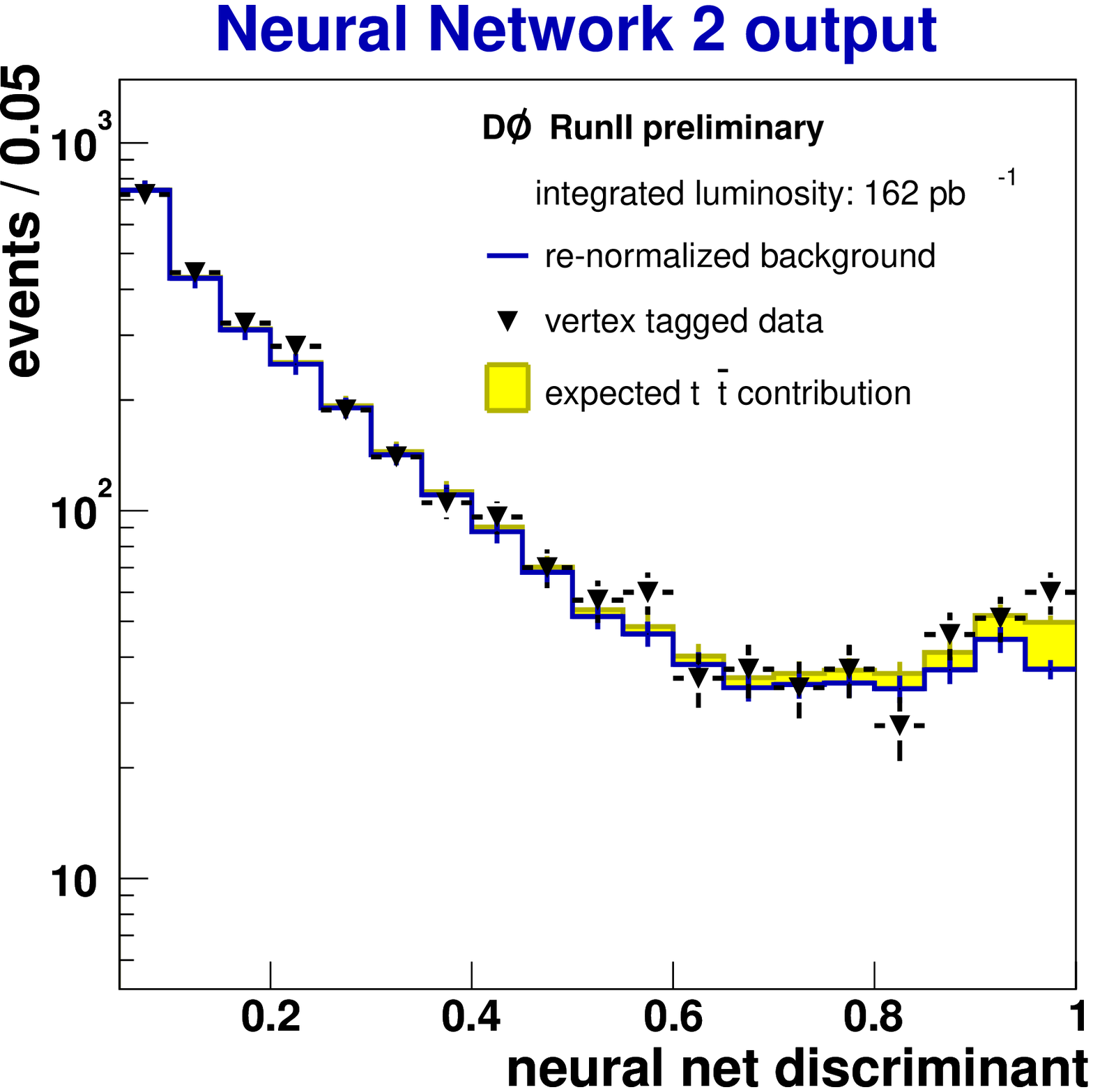}
  \caption{\it
    CDF and \Dzero\ $\ttbar$ cross-section measurements in the
    hadronic channel.  Left: number of b tagged jets in the CDF event sample
    as a function of jet multiplicity.
    Right: output of the final NN in the \Dzero\ analysis.
    \label{alljetsxs.fig} }
\end{figure}

To separate $\ttbar$ events in the hadronic channel 
from the large multijet background, both tight 
kinematic cuts and b tagging information are applied.
CDF selects events with 6 to 8 jets and no isolated leptons and
applies kinematic cuts.
In the distribution of the 
number of b tagged jets as a function of jet multiplicity
(see figure~\ref{alljetsxs.fig})
the $\ttbar$ cross-section is then measured to 
be\cite{bib-CDFalljetsxs}
\begin{equation}
\sigmattbar = \left(7.8 \pm 2.5\stat
                        \pandm{4.7}{2.3}\syst\right)\pb
\end{equation}
in $165\invpb$.

\Dzero\ selects events with 6 or more jets, of which exactly one 
is required to be b tagged.
A chain of NNs feeding into each other is used,
and the $\ttbar$ cross-section is determined from the excess of 
events after a cut on the last NN output over
background.
A data sample of $162\invpb$ yields\cite{bib-Dzeroalljetsxs}
\begin{equation}
\sigmattbar = \left(7.7 \pandm{3.4}{3.3}\stat
                        \pandm{4.7}{3.8}\syst
			\pm 0.5\lumi\right)\pb \ .
\end{equation}

\subsection{Events with $\W\to\taunu$ Decays}

CDF searches for $\ttbar$ events where one \W\ decays electronically
or muonically, while the other decays into a $\taunu$ final state
with a subsequent $\tau$ decay into hadron(s) and a 
neutrino\cite{bib-CDFbrtau}.
Events with an electron or muon with $\Et>20\GeV$, a 
tau lepton with $E>15\GeV$ and opposite charge, missing transverse energy
$\Etmiss>20\GeV$, at least two jets ($\Et(1)>25\GeV$,
$\Et(2)>15\GeV$), and $\Ht>205\GeV$ are selected.
The two events observed in $193.5\invpb$ are consistent with the 
Standard Model expectation, and a limit of
\begin{equation}
Br(\topquark\to\bquark\tau\nu) < 5.0\cdot Br_{\rm SM}(\topquark\to\bquark\tau\nu)
\end{equation}
is derived at 95\% confidence level.

\subsection{Summary of $\ttbar$ Cross-Section Measurements}

\begin{figure}[t]
  \vspace{5.5cm}
  \includegraphics{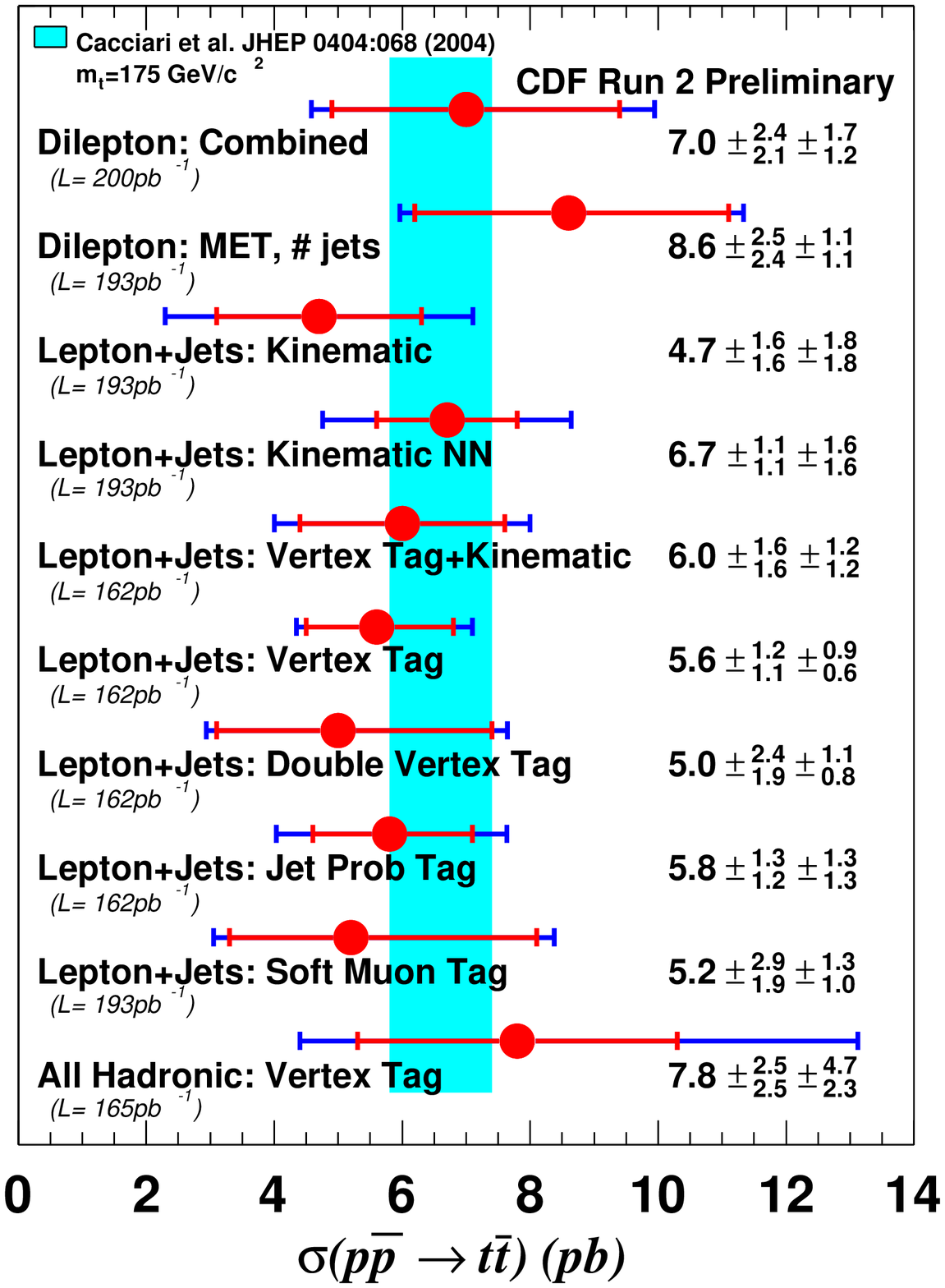}
  \includegraphics{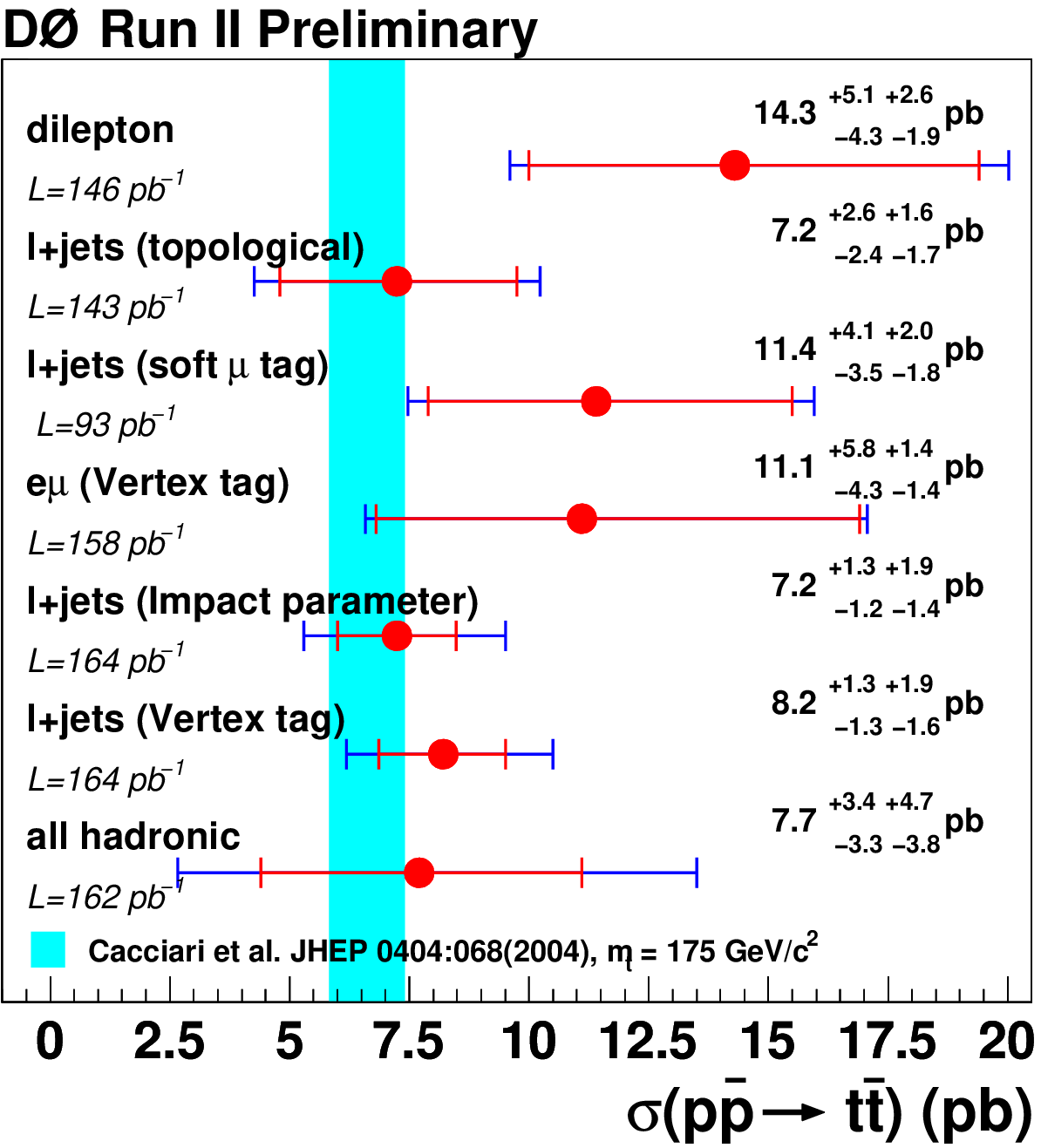}
  \caption{\it
    A summary of all Tevatron \runii\ $\ttbar$ cross-section
    measurements by CDF (left) and \Dzero\ (right).
    \label{allxs.fig} }
\end{figure}

The $\ttbar$ cross-section measurements at Tevatron \runii\ are
summarized in figure~\ref{allxs.fig}.  
The measurements in all decay channels and by both CDF and 
\Dzero\ are mutually consistent and consistent with the prediction 
of the Standard Model.

\section{Further \boldmath$\ttbar$\unboldmath{} Measurements}

\label{furtherttbar.sec}

It is conceivable that physics beyond the Standard Model does not 
change the total $\ttbar$ cross-section, but either only affects
differential cross-sections or top quark decays.

\subsection{Searches for New Physics in $\ttbar$ Production}

\begin{figure}[t]
  \vspace{4.5cm}
  \includegraphics{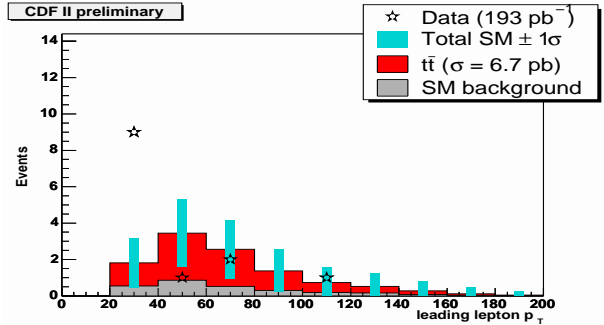}
  \caption{\it
    CDF search for anomalous dilepton event properties.
    The $\pt$ spectrum of the leading charged lepton in dilepton 
    $\ttbar$ events at CDF.
    \label{anomalouskinematics.fig} }
\end{figure}

In a model independent analysis, 
CDF have searched for anomalous kinematic properties in their
dilepton $\ttbar$ sample\cite{bib-CDFanomalouskinematics}.
Four kinematic distributions where new physics signatures are 
expected to be likely to be seen were chosen a priori.
While one of them, the leading lepton $\pt$ spectrum shown in 
figure~\ref{anomalouskinematics.fig}, shows an excess at low
transverse momenta in $193\invpb$, 
the other distributions agree with the expectation.
The overall compatibility with the Standard Model prediction has
been computed to be in the $1.0-4.5\%$ range.

\begin{figure}[t]
  \vspace{3.7cm}
  \includegraphics{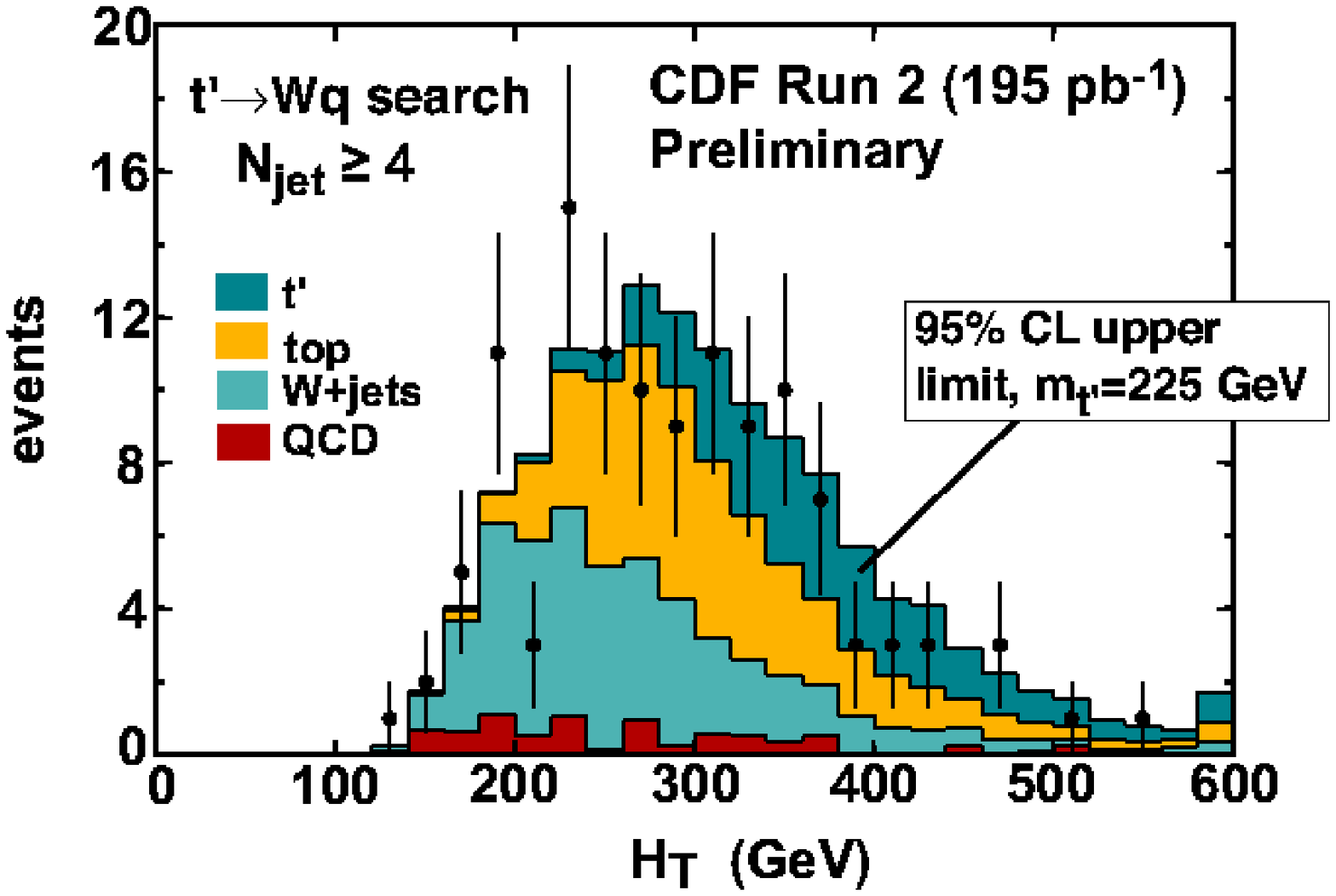}
  \includegraphics{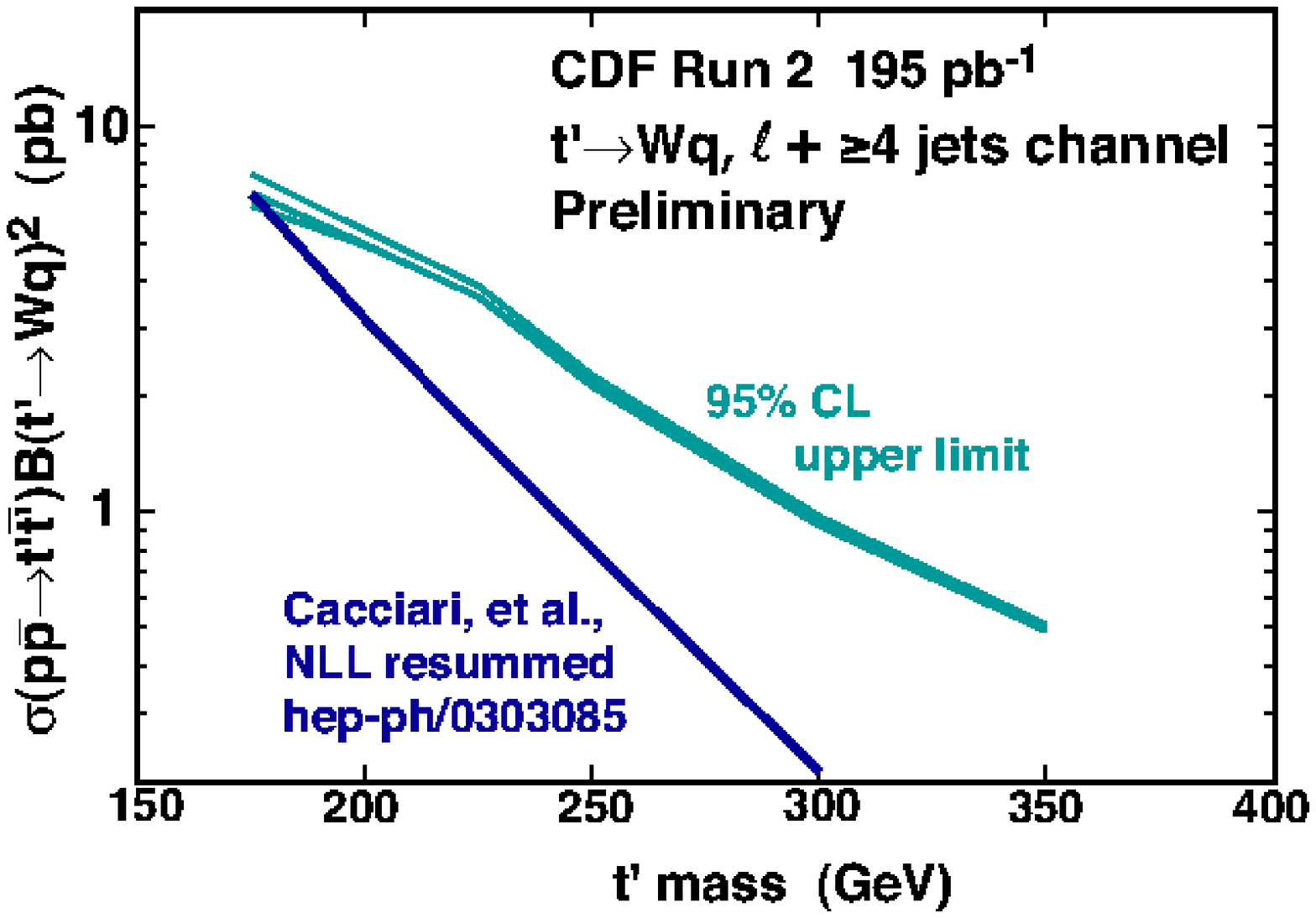}
  \caption{\it
    CDF $\tprime$ search.
    The $\Ht$ distribution in W+4jets events at CDF with a 
    topological selection is shown with a fit that includes
    a component from hypothetical $\tprime$ quark production (left).
    The resulting $\tprime$ cross-section limits as a function 
    of assumed $\tprime$ mass are also shown (right).
    \label{tprimesearch.fig} }
\end{figure}

CDF searches explicitly for production of fourth generation 
quarks\cite{bib-CDFtprimesearch}.
If these $\tprime$ quarks are heavier than the top quark, an excess
of events at large $\Ht$ is expected.
From a fit to the measured $\Ht$ distribution, which is consistent
with the Standard Model expectation, upper limits on the 
cross-section of $\tprime\overline{\tprime}$ events can be placed, see
figure~\ref{tprimesearch.fig}.

\subsection{Searches for New Physics in Top Quark Decays}

\begin{figure}[t]
  \vspace{3.0cm}
  \includegraphics{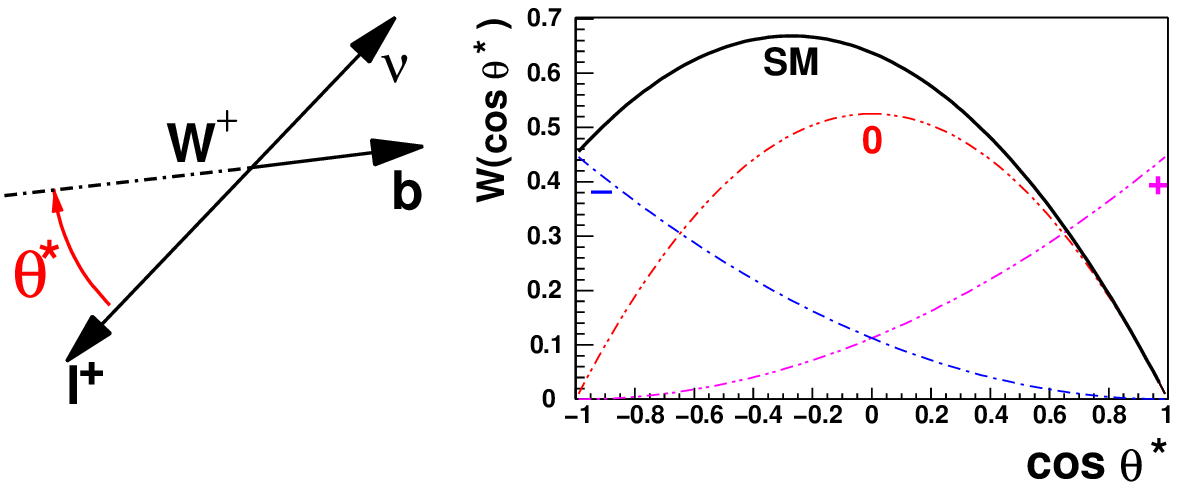}
  \includegraphics{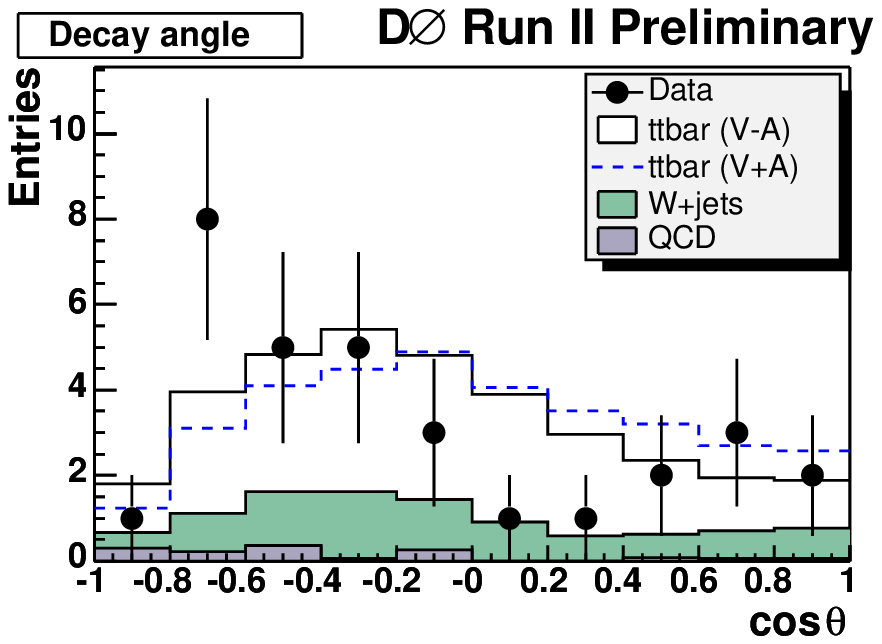}
  \caption{\it
    \W\ helicity measurements.
    The definition of the angle $\theta^*$ in the $\W$ rest frame
    (left) and the expected $\cos\theta^*$ distribution (middle),
    compared with the measured $\cos\theta^*$ distribution in 
    b tagged lepton+jets events from $159-169\invpb$
    at \Dzero\ (right).
    \label{costhetastar.fig} }
\end{figure}

The Standard Model predicts the fractions of longitudinal
and left-handed $\W$ bosons from top decay to be 
$F_0 = 1/(1+2\mW^2/\mtop^2) \approx 0.7$ and $F_- = 1-F_0$, while the 
fraction $F_+$ of right-handed W bosons is essentially zero because
the bottom quarks from top quark decay are left-handed due to
the large mass difference between top and bottom quarks.
The predicted distribution of the decay angle
$\theta^*$ in the $\W$ rest frame is shown in
figure~\ref{costhetastar.fig}. 
From measurements of this distribution (or quantities that depend
on $\cos\theta^*$), one can either search for non-zero constributions
from right-handed $\W$ bosons ($F_+>0$) or, assuming $F_+=0$, 
for deviations from the predicted ratio $F_0/F_-$.

CDF have measured the fraction $F_0$ from the charged lepton
$\pt$ spectrum (using $200\invpb$) to be\cite{bib-CDFleptonptWhel}
\begin{equation}
F_0 = 0.27 \pandm{0.35}{0.24}\ ,
\end{equation}
and from explicit reconstruction of the value of 
$\cos\theta^*$ (using $162\invpb$)\cite{bib-CDFcthsWhel} to be
\begin{equation}
F_0 = 0.89 \pandm{0.30}{0.34}\stat \pm 0.17\syst\ ,
\end{equation}
respectively -- 
to be compared with the \Dzero\ \runi\ value of\cite{bib-DzeroWhelRunI}
$F_0 = 0.56 \pm 0.31$.
All of these values are consistent with the Standard Model
expectation.

The two \Dzero\ measurements both 
use explicit $\cos\theta^*$ reconstruction in an event sample
obtained with a topological\cite{bib-DzerotopoWhel} selection 
or using b tagging\cite{bib-DzerobtagWhel} in
$159-169\invpb$.
The $\cos\theta^*$ distribution from the b tagging analysis
is shown in figure~\ref{costhetastar.fig}.
Both analyses each yield a limit of 
\begin{equation}
F_+<0.24\ {\rm at\ 90\%\ confidence\ level,}
\end{equation}
to be compared with 
the CDF \runi\ exclusion limit of $F_+<0.18$ at 95\% confidence 
level\cite{bib-CDFWhelRunI}.

\begin{figure}[t]
  \vspace{5.1cm}
  \includegraphics{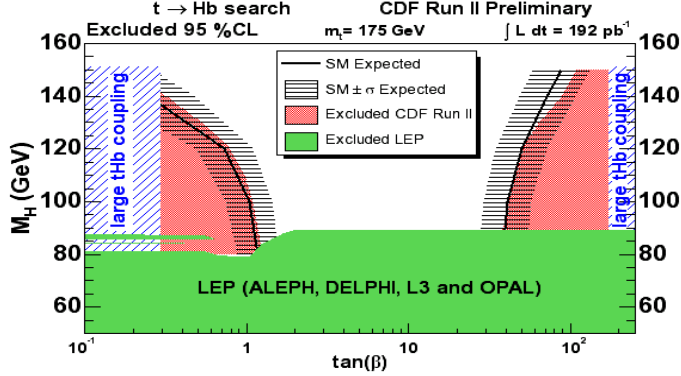}
  \caption{\it
    CDF charged Higgs search.
    Charged Higgs limits in the $m_{H^\pm}$ 
    vs.\ $\tan\beta$ plane.  The pink regions have been excluded,
    and the black lines with error bars indicate the expected limit.
    \label{chargedhiggslimits.fig} }
\end{figure}

In supersymmetric models with $m_{H^\pm}<\mtop$, the top quark may 
decay into a charged Higgs and a bottom quark.
Depending on the values of $\tan\beta$ and 
$m_{H^\pm}$, one expects the following changes in the observed $\ttbar$ event
topologies\cite{bib-topquarkphysics}:
\begin{list}{$\bullet$}{\setlength{\itemsep}{0ex}
                        \setlength{\parsep}{0ex}
                        \setlength{\topsep}{0ex}}
\item
an excess of $\tau$ decays due to $H^+\to\tau^+\nu$ decays for
large $\tan\beta$,
\item
an excess of hadronic top decays due to $H^+\to c\overline{s}$
decays for small $\tan\beta$ and small $m_{H^\pm}$, or 
\item
$\ttbar$ events with two extra b jets from $H^+\to \W^+ \bbbar$ decays
for small $\tan\beta$ and large $m_{H^\pm}$.
\end{list}
The CDF collaboration has therefore taken their measurements of the 
$\ttbar$ cross-section in the dilepton and lepton+jets channels as
well as their limit on $\ttbar\to\ell+\tau$ events\cite{bib-CDFchargedhiggs}
to place limits on $\topquark\to\Hp\bquark$ decays in the $m_{H^\pm}$ 
vs.\ $\tan\beta$ plane, as shown in
figure~\ref{chargedhiggslimits.fig}.

Both CDF and \Dzero\ have compared their $\ttbar$ cross-section
measurements obtained with different numbers of b tagged jets to determine
the branching ratio
$Br(\topquark\to\W\bquark)/Br(\topquark\to\W\q)$, 
where $\q$ denotes any down-type quark.
The results are\cite{bib-Brtbtq}
\begin{equation}
\begin{array}{lr@{\ }l}
1.11 \pandm{0.21}{0.19} & {\rm CDF}, & 162\invpb, \\
0.65 \pandm{0.34}{0.30}\stat \pandm{0.17}{0.12}\syst & {\rm\Dzero}, & 158-169\invpb,\
and\\
0.70 \pandm{0.27}{0.24}\stat \pandm{0.11}{0.10}\syst & {\rm\Dzero}, & 158-169\invpb,\
\end{array}
\end{equation}
where the first \Dzero\ result has been obtained with impact parameter
b tagging and the second with secondary vertex b tagging.
They show no sign of a deviation from the Standard Model expectation 
close to 1.
It should be noted that this quantity does not constrain the value
of $|\Vtb|^2$ in models where top quark decays into quarks from
more than three quark generations are allowed.

In summary, from measurements of $\ttbar$ production, there is
currently no sign of physics beyond the Standard Model.

\section{Search for Single Top Quark Production}

\label{singletop.sec}

The production cross-section for single top quarks is proportional
to $|\Vtb|^2$. 
Also, any differences to the Standard Model prediction could provide
hints for new physics.

In their searches for single top quark production, the Tevatron 
experiments concentrate on s-channel and t-channel production with 
expected cross-sections of $0.9\pb$ and $2.0\pb$, respectively,
see figure~\ref{feyndiag.fig}.

\begin{figure}[t]
  \vspace{3.6cm}
  \includegraphics{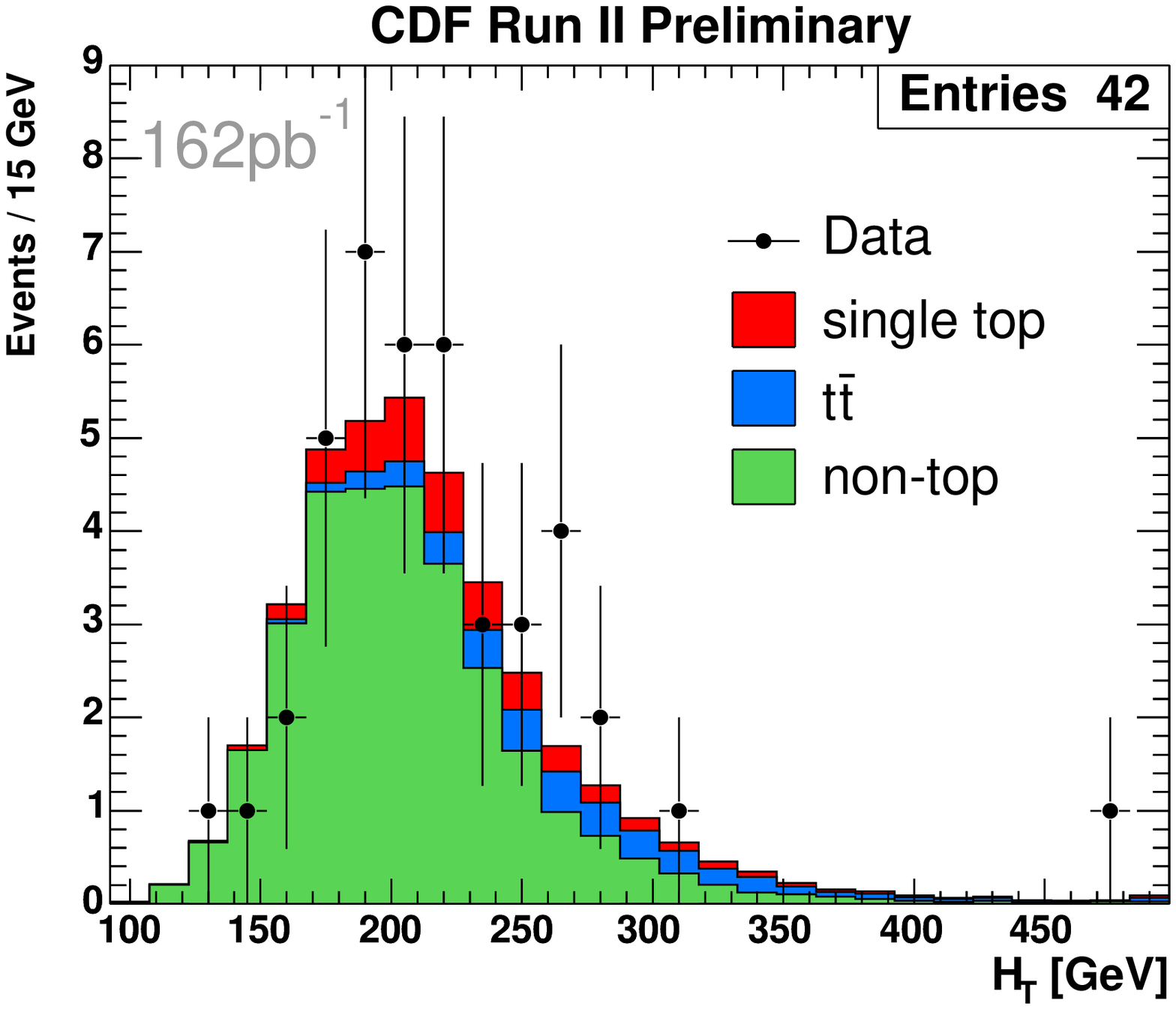}
  \includegraphics{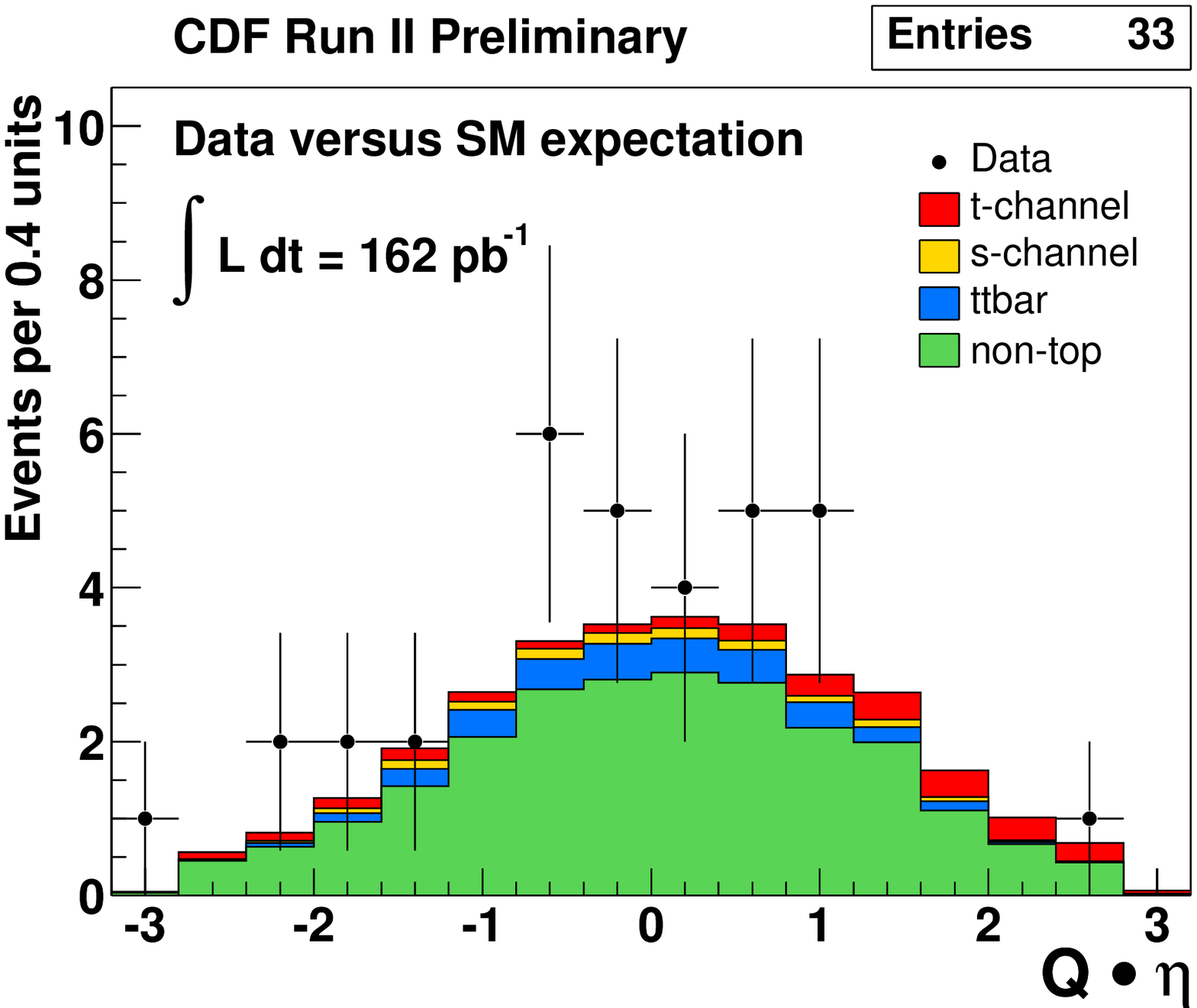}
  \caption{\it
    CDF single top search:
    The $\Ht$ distribution for single top quark candidate events at
    CDF (left) and the lepton charge signed $\eta$ distribution of 
    the b jet (right) together with the Standard Model expectations
    for non-top and $\ttbar$ backgrounds and single top signal (the
    contributions from s-channel and t-channel events are shown
    separately in the right plot).
    \label{CDFsingletop.fig} }
\end{figure}

Both CDF and \Dzero\ select events with an energetic isolated
charged lepton, missing transverse energy, and exactly 2 (CDF)
or 2--4 (\Dzero) jets out of which at least one must be b tagged.
CDF then selects events with a reconstructed top quark mass between 
$140\GeV$ and $210\GeV$, while \Dzero\ requires $\Ht>150\GeV$.
As shown in figure~\ref{CDFsingletop.fig},
single top events can be found at intermediate values of $\Ht$, and 
s-channel and t-channel events can be disentangled using the
lepton charge signed distribution of the pseudorapidity of the
identified b jet.
With the current data sets, sensitivity for Standard Model single top 
quark production has not yet been reached.
No significant excess of events has been observed, and the following 
95\% confidence level 
limits have been placed on the single top quark 
cross-section\cite{bib-singletop}:
\begin{equation}
\begin{array}{c|ccc}
{\rm experiment} & {\rm s-channel} & {\rm t-channel} & {\rm s\!+\!t-channel} \\
\hline
{\rm CDF}        & 13.6\pb         & 10.1\pb         & 17.8\pb \\
{\rm\Dzero}      & 19\pb           & 25\pb           & 23\pb\\
\end{array}
\end{equation}

With more data being taken and analysed and refined methods being 
developed, sensitivity for Standard Model single top quark
production is within reach for Tevatron \runii.

\section{Conclusions}

\label{conclusions.sec}

The current status of top quark measurements at the Tevatron
experiments CDF and \Dzero\ has been summarized, with the 
exception of the results for the top quark mass which are covered
in a separate article\cite{bib-velev}.

A wealth of measurements of the total $\ttbar$ production
cross-section are available from Tevatron \runii.
Measurements have been performed for dilepton,
lepton+jets, all-hadronic events, and events with top quark 
decays involving $\tau$ leptons.
They all yield results that are both mutually
consistent and in agreement with the Standard Model prediction.

The event samples have been further interpreted by looking for 
non-Standard Model $\ttbar$ production mechanisms and
top quark decays.
No signs for physics beyond the Standard Model have been found so 
far, supporting the interpretation of the signal as $\ttbar$
production via QCD and top quark decay to $\W\bquark$ final states.

In the search for single (electroweak) production of top quarks,
the sensitivity of the experiments has been improved over \runi.
Even for Standard Model single top quark production,
a significant cross-section measurement at the Tevatron is within reach
in the near future.

\section{Acknowledgements}

The author would like to thank the organizers for a very interesting 
and enjoyable conference.

\end{document}